# Light-matter coupling and non-equilibrium dynamics of exchange-split trions in monolayer WS$_2$

Jonas Zipfel,[1] Koloman Wagner,[1] Jonas D. Ziegler,[1] Takashi Taniguchi,[2] Kenji Watanabe,[3] Marina A. Semina,[4] Alexey Chernikov [1]

[1] *Department of Physics, University of Regensburg, Regensburg, Germany*
[2] *International Center for Materials Nanoarchitectonics, National Institute for Materials Science, Tsukuba, Japan*
[3] *Research Center for Functional Materials, National Institute for Materials Science, Tsukuba, Japan*
[3] *Ioffe Institute, Saint Petersburg, Russian Federation*

Monolayers of transition metal dichalcogenides present an intriguing platform to investigate the interplay of excitonic complexes in two dimensional semiconductors. Here, we use optical spectroscopy to study light-matter coupling and non-equilibrium relaxation dynamics of three-particle exciton states, commonly known as trions. We identify the consequences of the exchange interaction for the trion finestructure in tungsten-based monolayer materials from variational calculations and experimentally determine the resulting characteristic differences in their oscillator strength. It allows us to quantitatively extract trion populations from time-resolved photoluminescence measurements and monitor their dynamics after off-resonant optical injection. At liquid helium temperature we observe pronounced non-equilibrium distribution of the trions during their lifetime with comparatively slow equilibration that occurs on time-scales up to several 100's of ps. In addition, we find an intriguing regime of population inversion at lowest excitation densities that builds up and is maintained during 10's of picoseconds. With increased lattice temperature the equilibrium is established more rapidly and the inversion disappears, highlighting the role of thermal activation for efficient scattering between exchange-split trions.

## I. INTRODUCTION

Single layers of transition metal dichalcogenides (TMDCs) offer an appealing platform to investigate many-body interactions of charge carriers in condensed matter. The two-dimensional (2D) quantum confinement effects[1, 2] and reduced dielectric screening from the environment[3-6], lead to a very strong Coulomb interaction in these systems, resulting in the formation of optically accessible, tightly bound excitonic states[7, 8]. Interestingly, the basic properties of these quasiparticles are generally reminiscent of concepts known from atomic physics and chemistry, with excitons in inorganic semiconductors often described in a hydrogen-like Wannier-Mott picture[9-13]. The resemblance to a hydrogen atom includes excitonic Rydberg series[10, 12] and the formation of more complex states involving multiple charge carriers. In TMDC monolayers, exciton quasiparticles containing up to five charge carriers[14-18] are found to be thermodynamically stable at experimentally accessible conditions.

Of particular interest in this context is the scenario of excitons being embedded in a sea of free carriers that can lead to formation of charged, ion-like states commonly known as trions[19-24]. The combination of room-temperature stability, efficient light-matter coupling as well as electrical, optical, and environmental tunability provides broad opportunities to explore trion complexes in TMDCs both from fundamental and technological perspectives. However, trions are usually created by optical injection of electron-hole pairs in the presence of free carrier doping[22, 25, 26]. As a consequence, trion populations are expected to be initially far from the thermodynamic equilibrium of the crystal and subsequently relax towards thermal distribution. In addition, the multi-valley nature of TMDC materials[27-29] yields a manifold of trion states giving rise to the trion finestructure[30-33] that is associated with free charge carriers being in different valleys. Together, this can lead to complex scenarios involving non-equilibrium dynamics determined by the properties and the interplay of multiple trion states involved.

The aim of this work is to explore non-equilibrium conditions and relaxation dynamics of trions in two-dimensional TMDCs. For this purpose, we study WS$_2$ monolayers, encapsulated with high-quality hexagonal boron nitride (hBN) to mitigate disorder[34-36] and allow for accurate quantitative observations. While a vast manifold of excitonic states is accessible in this system, including spin- and momentum-dark excitons[37-41], we focus on the bright doublet of the negatively charged trion. It originates from two different electron spin-valley configurations and is energetically split due to exchange interaction[32, 42]. Importantly, these states can be addressed with both emission- and absorption-type techniques. As a consequence, we can directly access light-matter coupling strength in direct comparison with the theoretical calculations of the trion finestructure as well as extract relative trion populations from the luminescence signals. In addition, the majority of hBN-encapsulated samples tend to exhibit very weak negative doping at low temperatures. That allows us to employ large ensemble statistics for quantitative evaluation but also to study the monolayers in the ultra-low carrier density regime where the trion and Fermi-polaron descriptions[43-45] are expected to merge.

In the experiments we demonstrate a clear, characteristic difference in the oscillator strength between two different trion configurations. From theoretical calculations it is identified to stem from the electron-electron exchange interaction being intimately coupled to the energy splitting of the doublet. Using these results we then extract the time-resolved distribution of the trion populations within the finestructure from ultrafast photoluminescence (PL) spectroscopy. We demonstrate that after pulsed, non-resonant optical excitation a highly non-equilibrium distribution is established that persists for several 100's of ps at liquid helium temperature. In this regime, the trions are essentially out of thermal equilibrium with the crystal lattice during their lifetime on the order of 100 ps. Interestingly, at lowest excitation densities the population ratio is initially inverted following the excitation, with the higher lying trion state being more occupied than the energetically favourable one. The inversion is maintained for about 50 ps. At elevated temperatures trion equilibrate much faster and the population is not observed, implying efficient, thermally activated scattering between exchange-split states.

## II. EXPERIMENTAL DETAILS

The investigated TMDC samples are monolayers of commercially available $WS_2$ crystals, fabricated by mechanical exfoliation and subsequent all-dry viscoelastic stamping[46]. The flakes are encapsulated between thin layers of high-quality hexagonal boron nitride (hBN) by alternating the stamping of individual layers onto a preheated (either 70 or 100 °C) $SiO_2$/Si substrate. After stamping the first hBN-layer the sample is annealed in high vacuum at 150°C for 3 to 4 hours. No further heat treatment is applied after the transfer of the $WS_2$ flake. Altogether, we typically obtain hBN-encapsulated monolayers with narrow spectral lines across areas of a few 100's of µm², albeit the total area of the heterostructure is usually larger. Successful encapsulation suppresses the effects of non-homogeneous surroundings that can otherwise lead to additional spectral broadening[34-36, 47], inhibiting clear observation of the trion finestructure. The samples are then placed inside an optical microscopy cryostat and cooled down to liquid helium temperature of about 5 K that also reduces homogeneous broadening from exciton-phonon scattering.

Linear optical response of both trion and exciton states is then studied in reflectance geometry using a spectrally broad tungsten-halogen whitelight source. The incident light is spatially filtered by a pinhole and then focused onto the sample by a 40x glass-corrected microscope objective to a spot with a full-width-at-half-maximum (FWHM) of about 2.5 µm. The reflected signal is detected using a CCD camera coupled to a grating spectrometer. A systematic scanning of large sample areas with 0.5 µm steps is employed for a quantitative evaluation of representative statistics.

Time-resolved photoluminescence studies are performed using a pulsed Ti:sapphire laser source with the photon energy tuned to 2.43 eV through second-harmonic generation of the fundamental mode. For the studied $WS_2$ monolayer samples it corresponds to off-resonant excitation conditions. The beam is focused to a spot size with 1.5 µm FWHM. The resulting PL is guided through a spectrometer and detected by a streak camera with a nominal time resolution used in experiments on the order of 4 to 5 ps.

## III. RESULTS AND DISCUSSION

**Oscillator strength analysis of the trion finestructure**

Steady state optical response of a hBN-encapsulated $WS_2$ monolayer at 5 K is illustrated in FIG. 1 in the range of the fundamental optical transition. FIG. 1 (a) shows a representative reflectance contrast spectrum, obtained by collecting the reflectance both from the sample

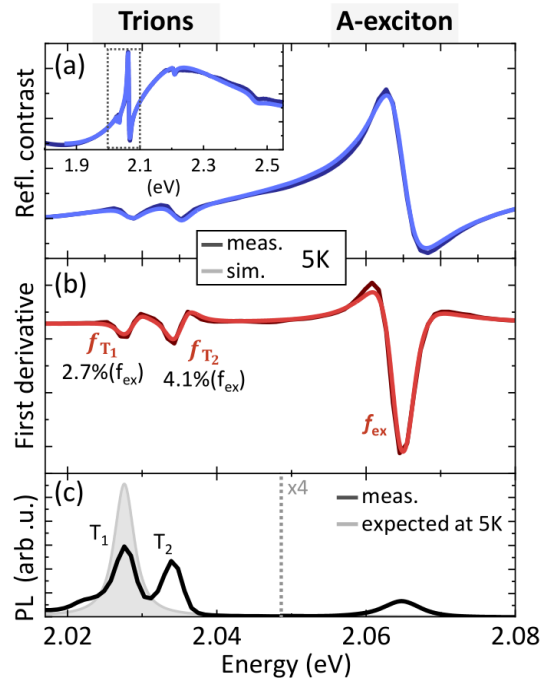

FIG. 1 (a) Reflectance contrast spectrum of hBN-encapsulated $WS_2$ at 5 K in the range of fundamental A-exciton transitions and weak electron doping, including a simulated curve from multi-Lorentzian parameterization of the dielectric function. The inset presents the spectrum in broad spectral range. (b) Corresponding first derivative of the reflectance contrast. The trion oscillator strengths $f_{T_1}$ and $f_{T_2}$ extracted from the simulated response are denoted as fractions of the A-exciton oscillator strength $f_{ex}$. (c) Time integrated photoluminescence PL detected after pulsed excitation. The neutral exciton resonance is enlarged for better visibility. The shaded area indicates the expected PL response assuming a thermal equilibrium distribution at 5 K given by Boltzmann statistics.

($R_S$) and the SiO$_2$/Si substrate as a reference ($R_{Ref}$) according to $R = (R_S - R_{Ref})/(R_{Ref} - R_{BG})$. The main resonance stems from the neutral A-exciton ground state and the energetically lower doublet corresponds to negatively charged, bright trions[24, 32, 42]. For an overview, a full spectrum in a broader spectral range is presented in the inset. FIG. 1 (b) shows the corresponding first derivative. Further included are simulated spectra using a model dielectric function $\varepsilon(E)$. It is composed of multiple Lorentzians of the form $f_j/(E_j^2 - E^2(\omega) - iE(\omega)\Gamma_j)$ for each exciton resonance $j$ with transition energy $E_j$, non-radiative linewidth $E_j$, and oscillator strength $f_j$, as well as including an offset $\varepsilon_0$ in the real part. Reflectance contrast spectra are then calculated via the transfer matrix approach[48] with the parameters of $\varepsilon(E)$ and the hBN-layer thickness adjusted to match the measured response assuming normal incidence conditions. We note that we checked for the influence of large angles also in the experiment and found them to be minor, likely related to a comparatively low numerical aperture of the spectrometer optics. This procedure allows us to quantitatively extract contributions for each individual excitonic transition to the linear response.

Here, the energies of the trion and exciton resonances $E_{ex} = 2.064$ meV, $E_{T_1} = 2.028$ meV and $E_{T_2} = 2.0345$ meV as well as the non-radiative broadening of $\Gamma_{ex} = 1.8$ meV and $\Gamma_{T_1} = \Gamma_{T_2} = 2.5$ meV are typical for hBN-encapsulated WS$_2$ monolayers[15, 34, 35, 49]. Narrow linewidths in these samples are associated with a strongly suppressed inhomogeneous broadening from reduced environmental disorder[35, 36, 50] that allows us to clearly observe and quantify the properties of the trion doublet. The extracted total oscillator strength of the two trions $f_{T_{tot}} = f_{T_1} + f_{T_2}$ is as low as 7% of that of the A-exciton resonance, well within the regime where doping-induced renormalization effects from screening or Pauli-blocking can be largely neglected.[8, 51, 52] It also provides a direct way to estimate the intrinsic doping level on the order of several 10$^{11}$ cm$^{-2}$, comparing it to scaling in gate-tunable devices based on WS$_2$[51].

The corresponding time-integrated PL spectrum under pulsed excitation is presented in FIG. 1(c). Here, a weak signal from the neutral A-exciton resonance is enhanced by a factor of 4 for better visibility. Interestingly, we observe light emission from all three peaks, despite rather sizable differences between the energies of the individual states in contrast to the thermal energy of about 0.4 meV at 5 K. In particular, both trions are separated by more than 6 meV, yet emit with very similar intensities. For comparison, the PL emission expected for thermal distribution of the trions at 5K is illustrated in FIG. 1(c) by the shaded area, obtained assuming Boltzmann distribution and using the oscillator strengths from reflectance. These observations demonstrate that the distribution of the photoexcited trions between the two states is far from equilibrium, indicating direct competition between relaxation and recombination timescales. In the following we investigate this non-equilibrium scenario in more detail. First, to infer population ratios from the detected emission we determine the light-matter coupling ratio of exchange-split trions in theory and experiment. This allows us to extract and monitor the dynamics of respective trion populations in time-resolved PL experiments, quantify their distribution, and identify the associated equilibration time scales.

The origin of the trion finestructure in WS$_2$ in the presence of free electrons is schematically illustrated in FIG. 2(a). On the left-hand side, the relevant bands at the K-points of the first Brillouin zone are depicted in single-particle picture, including only the upper valence band. The electron spin states are coupled to the valley index[27-29, 53] and denoted by different colors. For the purpose of illustration we consider the optically allowed transition only at the K+ valley as it will equally apply for the K- valley by changing the spin and valley signs. In this picture, the excitons are created by absorbing a photon and promoting an electron from the valence band to the *upper* conduction band. Considering the conduction band splitting on the order of 10's of meV, free electrons from residual doping occupy only the *lower* conduction bands in K+ and K- valleys if their density is not too high (typically, below several 10$^{12}$ cm$^{-2}$). Combining an exciton and an electron to form a composite trion state then results in two possible configurations, indicated by shaded areas in FIG. 2(a). One of the states includes both electrons in the same valley with antiparallel spins and is therefore commonly referred to as a *singlet* trion. In the other configuration the two electrons are located in different valleys with parallel spins and the state is thus commonly referred to as a *triplet* trion[32, 33, 42]. Here we note that due to the distinct valley configurations of the extra electron, these states can be also denoted as *intra-* and *inter*valley trions, respectively[54, 55].

A direct consequence of different electron spin configurations is the impact of short-range electron-electron exchange interaction on the trion properties. Most prominently, it results in the energy splitting of the two trion states as previously demonstrated[3, 32, 56] and schematically illustrated in the quasiparticle picture on the right-hand-side in FIG. 2(a). To investigate the impact of short-range exchange interaction on the oscillator strength in relation to the finestructure energy splitting, we first approximate it as a δ-potential in real space:

$$V_\pm^{ee} = U_\pm \delta(\vec{\rho}_{e1} - \vec{\rho}_{e2}). \quad (1)$$

Here, $\vec{\rho}_{e1}$ and $\vec{\rho}_{e2}$ are the relative coordinates of the two electrons with respect to the hole at origin and $U_\pm$ is the short-range interaction constant with $U_+ = -U_-$. The signs "+" and "-" indicate the location of the additional electron in either K+ or K- valley, respectively. We emphasize that the absolute sign of the exchange-interaction constant is *not known a priori* and thus can not be assigned to a *specific* configuration at this stage. As a consequence, even if a particular assignment of the trion peaks to singlet and triplet has been previously proposed, this analytical formalism does not allow to explicitly determine the particular configuration in the material. Throughout the manuscript we thus refer to the experimentally observed trion states by their phenomenological indices "1" and "2" and use "+" and "-" only in the theoretical notation following the above definition.

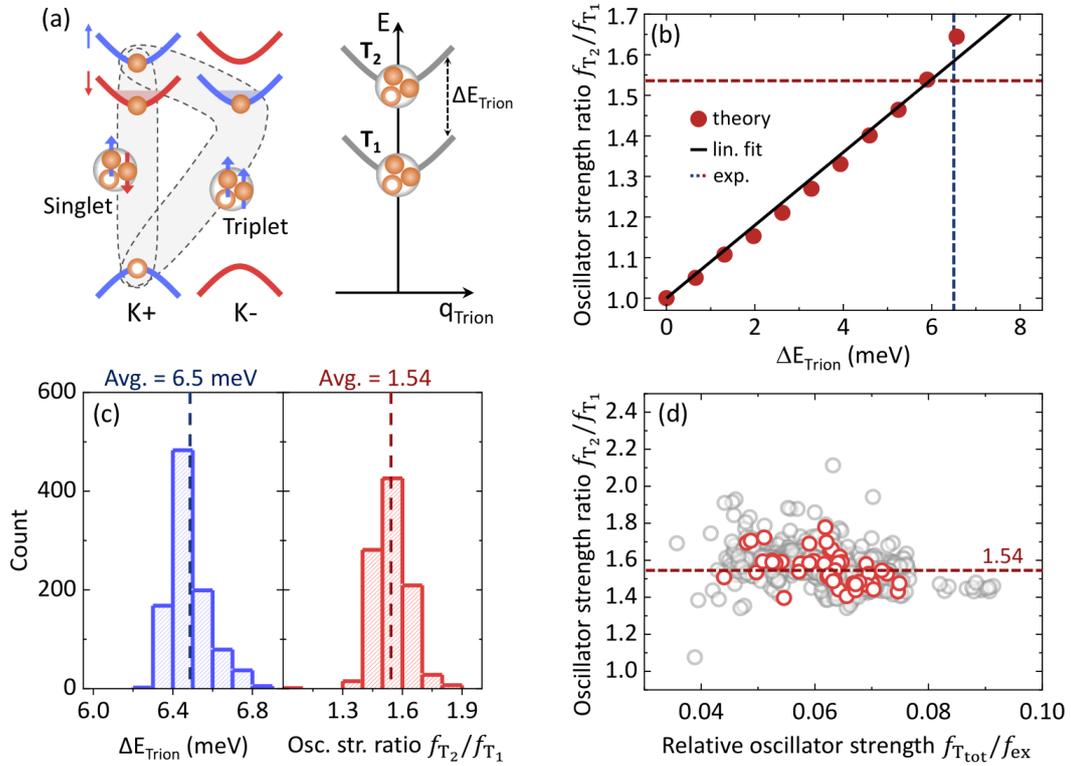

FIG. 2 (a, left panel) Schematic illustration of two negatively charged, bright trion configurations in a WS$_2$ single particle picture: intravalley singlets (total electron spin zero) or intervalley triplets (total electron spin one). (Right panel) Corresponding trion states in quasiparticle picture, energetically split due to short range exchange interaction. (b) Theoretical calculations of the oscillator strength ratio $f_{T_2}/f_{T_1}$ of the trion doublet as function of energy splitting. (c) Measured energy splitting (left) and oscillator strength ratio (right) obtained from a total of 976 individual reflectance contrast measurements across a hBN-encapsulated WS$_2$ monolayer sample, presented as histograms. (d) Trion oscillator strength ratio as function of the total oscillator strength relative to the neutral exciton. For better visibility every 24$^{th}$ data point is highlighted.

Importantly, regardless of the absolute sign of the constant $U_\pm$, it is possible to obtain both the size of the energy splitting $\Delta E_T = |\Delta E_+ - \Delta E_-|$ and the oscillator strength ratio related to that[32]. For this purpose we use a variational approach using the most simple trial function for the trion composed from two symmetrized ground state hydrogen-like electron wavefunctions[3, 32] of the form

$$\varphi_\pm(\rho_{e1}, \rho_{e2}) =$$
$$A_\pm(e^{-\rho_{e1}/a_{x,\pm} - \rho_{e2}/a_{tr,\pm}} + e^{-\rho_{e2}/a_{x,\pm} - \rho_{e1}/a_{tr,\pm}}), \quad (2.1)$$

with the normalization factor

$$A_\pm^{-2} = (2\pi)^2 \left( \frac{a_{x,\pm}^2 a_{tr,\pm}^2}{8} + \frac{2 a_{x,\pm}^4 a_{tr,\pm}^4}{(a_{x,\pm} + a_{tr,\pm})^4} \right), \quad (2.1)$$

where $a_x$ and $a_{tr}$ are the variational parameters corresponding to the Bohr constants of the two electrons relative to the hole. The calculation shows, that one of the variational parameters is close to the exciton Bohr radius ($a_x$), while the second one corresponds to the larger radius of the additional electron localized on the exciton ($a_{tr}$). In absence of the exchange interaction ($U_\pm = 0$) the values are $a_x$ = 13.5 Å and $a_{tr}$ = 40 Å. For finite $U_\pm$ we note that $a_{x,+}$ and $a_{x,-}$ (similarly $a_{tr,+}$ and $a_{tr,-}$) are close but not exactly the same for the two different electron configurations in singlet and triplet trions and slowly change with the short-range interaction strength.

The trial functions from EQ. (2.1) then enter an effective-mass Hamiltonian that includes the direct long-range Coulomb interaction in the 2D limit of the thin film model[5, 6] to account for non-uniform dielectric screening and the short-range exchange interaction from EQ. (1), see Ref [32] for details. Typical parameters used for hBN-encapsulated WS$_2$ are dielectric constant of hBN $\varepsilon_{hBN}$ = 4.5, screening length $r_0$ = 2.1 nm, and effective masses of the electron and hole $m_e = m_h = 0.32 m_0$ as fractions of the free electron mass $m_0$. The direct, long-range Coulomb interaction provides the largest contribution to the total trion binding energy and the exchange interaction determines the energy splitting between singlet and triplet trions. The value of the energy splitting calculated as a function of the trion Bohr parameters $a_x$ and $a_{tr}$ as well as the exchange interaction constant $U_\pm$ is

$$\Delta E_\pm = 2\pi A_\pm^2 U_\pm \frac{a_{x,\pm}^2 a_{tr,\pm}^2}{(a_{x,\pm} + a_{tr,\pm})^2}. \quad (3)$$

Thus, the energy degeneracy of the two trion states is lifted by the exchange coupling and is accompanied by the corresponding changes of the oscillator strength $f_T$ of the trions:

$$f_{T,\pm} = |M_r|^2 N_{e,\pm} (2\pi A_\pm)^2 (a_{x,\pm}^2 + a_{tr,\pm}^2)^2, \quad (4)$$

where $M_r$ is the dipole matrix element of the electron interband transition and $N_{e,\pm}$ is the electron density in the respective $K+$ or $K-$ valley. The latter is expected to be the same in absence of an out-of-plane magnetic field, i.e., $N_{e,+} = N_{e,-}$. We also note that for $a_x \ll a_{tr}$ the oscillator strength in EQ. (4) is roughly proportional to $a_{tr,\pm}^2$, i.e., to the squared trion radius. Evaluating the expressions (3) and (4) in accordance with previous works[32, 57-59], the ratio of oscillator strength is obtained as function of the energy splitting. The results are presented in FIG. 2(b) in the relevant range as function of the absolute splitting energy. The oscillator strength ratio is found to be roughly proportional to the energy splitting, visualized by the linear fit to the calculations as it was expected from theoretical estimations.

To obtain the ratio of trion oscillator strengths in experiment we employ systematic reflectance contrast scanning over a sample area of about 225 μm² grid (976 spectra in total, similar to that in FIG. 1 (a). We extract the underlying parameters of the trion resonances using transfer matrix method and Lorentzian parameterization as described above. The results for the measured trion splitting and oscillator strength ratio are presented in FIG. 2 (c) as histograms. We find the trion splitting to be very robust over the whole investigated area with an average value of $\Delta E_{Trion} = 6.5 \pm 0.2$ meV, in good agreement with literature values for individual representative measurements[15, 33, 42, 49]. The corresponding average trion oscillator strength ratio is $f_{T_2}/f_{T_1} = 1.54 \pm 0.12$. Importantly, it is independent from the fluctuations of doping in the studied low-density range, as shown by plotting the oscillator strength ratio as function of the total oscillator strength in FIG. 2 (d) that is proportional to doping. The spread of the data likely stems from the combined fitting error of the trion and exciton resonances.

Interestingly, the experimentally determined pair of values for the average energy splitting and oscillator strength ratio are found to be very well described by the predictions of the theoretical model (see FIG. 2 (b)). We can thus understand the measured, characteristic difference in the light-matter coupling of the trions as a direct consequence of the short-range exchange interaction between the electron constituents. In particular, it means that the state at higher energy ($T_2$) has a larger trion radius compared to the lower lying one ($T_1$). Moreover, we are now able to quantitatively extract distribution of the trion population from the relative strength of the PL signals that we use for the analysis of the time resolved data in the following.

**Population distribution dynamics**

To analyse trion distribution dynamics we turn to the temporal evolution of the PL response after pulsed optical excitation. FIG. 3 (a) shows a representative time- and spectrally resolved streak camera image in the spectral range of the trion doublet at 5 K. The incident energy density of the pump pulse is set to 0.8 μJ/cm², corresponding to a photoinjected electron-hole pair density on the order of $1 \times 10^{11}$ cm$^{-2}$ (obtained assuming an effective absorption of 4.5% at the excitation photon energy, estimated from linear response). To highlight the relative distribution of the

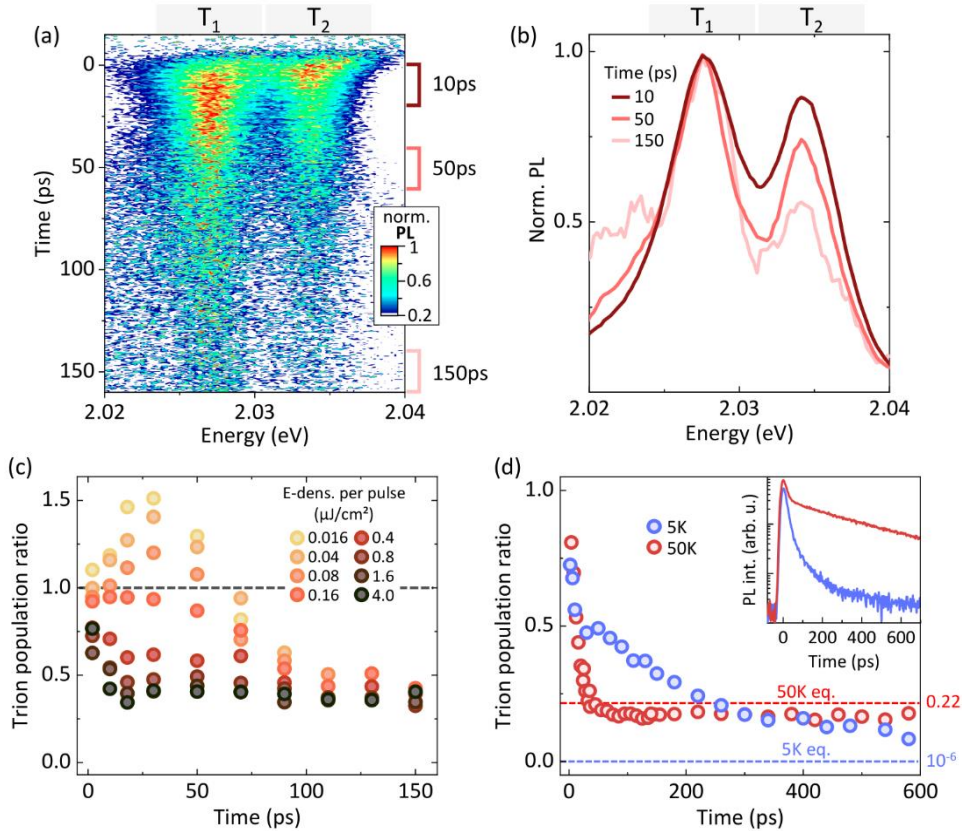

FIG. 3 (a) Normalized streak camera image of the time- and spectrally resolved trion PL response for an excitation density of 0.8 μJ/cm². (b) Normalized PL spectra for different times after excitation, integrated over intervals of Δt = 20 ps as indicated in (a). The data are slightly smoothed for better visibility. (c) Trion population ratio $T_2/T_1$ (PL ratio / oscillator ratio) as a function of time for different excitation densities. The shaded area indicates the regime of population inversion. (d) Trion population ratio for different lattice temperatures of 5 and 50 K for the 0.8 μJ/cm² excitation density. Solid lines indicate the population equilibrium given by a Boltzmann distribution 0.22, $10^{-6}$. The inset depicts the transients of the total trion PL.

luminescence intensity, the data is normalized at each individual time step. Representative spectra at 10, 50, and 150 ps, integrated over a time range of Δt = 20 ps are presented in FIG. 2(b). The data are smoothed by averaging over a ΔE = 1 meV interval and normalized to the $T_1$ trion intensity to emphasize intensity ratios. At these conditions, the emitted PL from the two trions is very similar directly after the excitation and slowly changes with time in favor of the low energy state.

For quantitative analysis, the emission intensities $I(T_1)$ and $I(T_2)$ of the two trion resonances are extracted from the respective areas of the PL spectra using multi-Lorentzian fit functions for each time frame (over 10 or 20 ps intervals). Population ratios $T_2/T_1$ are then obtained by dividing the PL intensity fractions $I(T_2)/I(T_1)$ by the oscillator strength ratio of 1.54. They are presented in FIG. 3(c) as function of time after the excitation over a range of excitation densities between 0.016 and 4 μJ/cm². Overall, we find a large fraction of the trion population to reside in the higher energy state following optical injection. We note that a similar finding has been previously reported in continuous wave PL of WSe$_2$[32], attributed to more efficient formation of the higher lying state proposed to stem from the energy separation to the neutral exciton being roughly in resonance with an optical phonon[54].

Interestingly, it seems that immediately after the excitation, within experimental resolution of a few ps, both trion states are almost equally populated for all excitation densities. At the lowest excitation densities that we applied, however, the trion distribution becomes initially inverted with time, meaning that the energetically higher trion is more strongly populated than the lower lying one. The inversion eventually decreases and disappears. At elevated pump densities the inversion is absent and the initial equilibration occurs directly after the excitation, reaching the ratio of about 0.5 after about 100 to 150 ps for all excitation densities. It is followed by a slower relaxation on a time scale of many 100's of ps, as illustrated in FIG. 3(c) for a selected excitation density of 0.8 μJ/cm². Notably, even after 600 ps as much as 10% of the trion population is found to reside in the higher-lying state. Altogether, during their lifetime on the order of 100 ps (obtained from the decay of the total emission intensity presented in the inset of FIG. 3(d)), the distribution of the trions between the exchange-split states is found to be far from thermal equilibrium conditions. The latter are estimated using Boltzmann distribution $T_2^{eq}/T_1^{eq} = \exp(\Delta E_{Trion}/k_B T)$ and added to FIG. 3 (d) as dashed lines (for T = 5 and 50 K).

The observations at 50 K, presented in FIG. 3 (d), are distinctly different. While the population ratio is roughly the same, close to unity after the excitation, it relaxes much faster in contrast to the 5 K results. The equilibrium is reached already after about 50 ps with a characteristic 1/e time constant of about 15 ps. The experimentally determined $T_2/T_1$ ratio after the equilibration is in reasonable quantitative agreement the one expected at lattice temperature of 50 K. We note, however, that particularly for longer delay times and increasingly weaker signals a third, long-lived resonance on the low energy side of the trion doublet has to be included in the fitting procedure. This gives rise to small systematic errors of the extracted trion population ratio at longer times and could lead to the measured values being slightly below the equilibrium prediction. Finally, population inversion is not observed at 50 K, even at lowest pump densities.

Altogether, these findings strongly indicate that the scattering channels between exchange-split trions are rather slow, e.g., compared to equilibration of excitons that should occur within a few to 10's of ps[60]. As a consequence, non-equilibrium distribution of the trions can be maintained for 100's of ps and even the population inversion of the trion doublet is stable for as long as 50 ps. Here we note that it is very likely that the trion center-of-mass momenta are thermally distributed and the trions are out of equilibrium only with respect to the finestructure splitting. In a simplified single-particle picture one may indeed expect weak coupling between singlet and triplet trions, since the flip of the electron spin would be required. However, in view of the trion being a composite many-body state, additional processes could be of importance. For example, a singlet trion from the interband exciton transition at the K+ valley could scatter to a triplet trion in the K- valley due to long-range exchange. In close analogy to neutral excitons, the excitonic part of the trion (core electron-hole pair) would then switch their valley index, while the additional electron would remain in the same state[61]. Alternatively, ionization of a trion followed by capture of a free electron from the opposite valley would effectively allow trion singlets to form triplets and vice versa. Both processes should exhibit pronounced thermal activation, qualitatively similar to the experimental observation. While the above speculations provide reasonable mechanisms, we emphasize, that the microscopic problem of the trion-trion scattering is generally very challenging and strongly motivates future efforts.

## IV. CONCLUSION

In summary, we have studied trion finestructure in hBN-encapsulated WS$_2$ monolayers under weak electron doping. We investigated the impact of the short-range exchange interaction on light-matter coupling and monitored non-equilibrium trion distributions and their dynamics after optical injection. From linear spectroscopy of the trion states and quantitative, statistical analysis we have demonstrated sizable differences in the oscillator strength within the trion doublet on the order of 50%. From theoretical calculations this observation is understood as a consequence of short-range exchange interaction, predicting a linear relationship between energy splitting and oscillator strength ratios. Following pulsed optical injection, we observed highly non-equilibrium distribution of the trion between exchange-split states. At liquid helium temperature and weak optical pumping population inversion emerged, lasting for as long as about 50 ps. The overall relaxation of the trion distribution is very slow at these conditions and the thermal equilibrium is not reached even at 600 ps after the excitation. In contrast, the equilibration is found to be much faster at an elevated temperature of 50 K, on a time-scale of 10's of ps. These findings suggest the presence of a temperature dependent bottleneck strongly limiting the coupling of the exchange-split trion states with very slow equilibration rates at low temperatures.

**ACKNOWLEDGEMENTS**

We thank Mikhail Glazov, Malte Selig, Andreas Knorr, Ermin Malic, and David Reichmann for helpful discussions. Financial support by the DFG via Emmy Noether Initiative (CH 1672/1-1) and SFB 1277 (project B05) is gratefully acknowledged. K.W. and T.T. acknowledge support from the Elemental Strategy Initiative, conducted by the MEXT, Japan, Grant Number JPMXP0112101001, JSPS KAKENHI Grant Numbers JP20H00354 and the CREST (JPMJCR15F3), JST, M.A.S. is gratefully acknowledging financial support from Russian Science Foundation (project no. 19-12-00273).

**DATA AVAILABILITY**

All relevant data are available from the corresponding author upon reasonable request.